# New classes of chiral topological nodes with non-contractible surface Fermi arcs in CoSi

Zhicheng Rao[1,2,*], Hang Li[1,2,*], Tiantian Zhang[1,2,*], Shangjie Tian[3,*], Chenghe Li[3], Binbin Fu[1,2], Cenyao Tang[1,2], Le Wang[1,2], Zhilin Li[1,4], Wenhui Fan[1,2], Jiajun Li[1,2], Yaobo Huang[5], Zhehong Liu[1,2], Youwen Long[1,2], Chen Fang[1], Hongming Weng[1], Youguo Shi[1], Hechang Lei[3,#], Yujie Sun[1,#], Tian Qian[1,#], and Hong Ding[1,2]

[1] *Beijing National Laboratory for Condensed Matter Physics and Institute of Physics, Chinese Academy of Sciences, Beijing 100190, China*

[2] *University of Chinese Academy of Sciences, Beijing 100049, China*

[3] *Department of Physics and Beijing Key Laboratory of Opto-electronic Functional Materials & Micro-nano Devices, Renmin University of China, Beijing 100872, China*

[4] *State key Laboratory for Artificial Microstructure and Mesoscopic Physic, Beijing Key Laboratory of Quantum Devices, Peking University, Beijing 100871, China*

[5] *Shanghai Synchrotron Radiation Facility, Shanghai Institute of Applied Physics, Chinese Academy of Sciences, Shanghai 201204, China*

[*] These authors contributed to this work equally.

Corresponding authors: tqian@iphy.ac.cn, yjsun@iphy.ac.cn, hlei@ruc.edu.cn

**In condensed matter systems, chiral topological nodes are robust band crossing points in momentum space that carry nonzero Chern numbers. The chirality is manifested by the presence of surface Fermi arcs connecting the projections of nodes with opposite Chern numbers[1]. In addition to the well-known Weyl nodes[1-4], theorists have proposed several other types of chiral topological nodes in condensed matter systems[5-10], but the direct experimental evidence of their existence is still lacking. Here, using angle-resolved photoemission spectroscopy, we reveal two types of new chiral nodes, namely the spin-1 nodes and charge-2 Dirac nodes, at the band crossing points near the Fermi level in CoSi, the projections of which on the (001) surface are connected by surface Fermi arcs. As these chiral nodes in CoSi are enforced at the Brillouin zone (BZ) center and corner by the crystalline symmetries, the surface Fermi arcs connecting their projections form a non-contractible path traversing the entire (001) surface BZ, in sharp contrast to pairs of Weyl points with small separation. Our work marks the first experimental observation of chiral nodes beyond the Weyl nodes both in the bulk and on the surface in condensed matter systems.**

In high-energy physics, there are three types of fermionic particles, i.e. Dirac, Weyl and Majorana fermions, predicted by the Standard Model based on the Poincaré group in the universe. Condensed matter systems can realize a variety of fermionic quasiparticles at topological nodes: robust band crossing points protected from being gapped by nontrivial band topology[1-18], which have or not have analogues in high-energy physics. Angle-resolved photoemission spectroscopy (ARPES) experiments have verified three types of topological nodes, i.e., Weyl nodes[19-23], Dirac nodes[24,25], and the three-fold degenerate nodes[26,27]. The Weyl nodes appear at linear crossings of two non-degenerate bands and carry nonzero Chern number $C = \pm1$. The nonzero Chern number makes Weyl nodes the pristine chiral topological nodes, and also necessitates the presence of exotic helicoid surface states having "Fermi arcs" as their equal-energy contour.

Aside from Weyl nodes, theorists have proposed more exotic chiral topological nodes, such as double-Weyl nodes with quadratic band crossing[5], three-component spin-1 nodes[6-8], charge-2 Dirac nodes and spin-3/2 Rarita-Schwinger-Weyl nodes with four-fold degeneracy[7-10], and double-spin-1 nodes with six-fold degeneracy[8-10]. Although first-principles calculations have proposed these unconventional chiral nodes in numerous materials[5-10], so far experimental evidence is still wanted. In this work, we have experimentally verified the chiral spin-1 nodes and charge-2 Dirac nodes in the transition metal silicide CoSi by investigating its electronic band structures of both bulk and surface states with systematic ARPES measurements.

Figure 1a illustrates the crystal structure of CoSi, in which each Co atom is bonded with six Si atoms and vice versa. They form a simple cubic structure of a lattice constant $a$ = 4.445 Å with space group $P2_13$ (No. 198). The corresponding BZ in Fig. 1b is also simple cubic with four high-symmetry momenta Γ, X, M, and R. First-principles calculations[8] have shown that the band structure has a three-fold degenerate point at Γ and a four-fold degenerate point at R near the Fermi level ($E_F$) (Fig. 1e), when spin-orbit coupling (SOC) is not considered. Similar degenerate points have also been identified in the calculated phonon spectra of CoSi (ref. 7). While considering SOC, the bands split due to the absence of inversion symmetry. However, the band splitting is of the order of meV because of weak SOC strength on the Co 3$d$ and Si 3$p$ orbitals[11]. Thus the SOC effects can be ignored in experimental work.

The quasiparticle excitations at the three-fold and four-fold degenerate points in CoSi are described as chiral spin-1 node and charge-2 Dirac node, respectively. They carry nonzero Chern numbers ±2 (refs. 7,8), which are distinct from the experimental verified three-fold degenerate nodes[26,27] and normal four-fold Dirac nodes[24,25]. According to the no-go theorem, the spin-1 and charge-2 points in CoSi must have opposite Chern numbers. It is expected to have two surface Fermi arcs connecting their projections on certain surfaces. Considering the band structure of CoSi, the Fermi arcs should lie in the gap between the bands #2 and #3. As shown in Fig. 1c,e, the bands #2 and #3 form the hole- and electron-like Fermi surfaces (FSs) that enclose the spin-1 and charge-2 points, respectively, leaving a large direct gap between them in the regions other than the FSs. This is important for observation of surface Fermi arcs at $E_F$.

As illustrated in Fig. 1d, the spin-1 and charge-2 points in CoSi are projected onto the surface BZ center and corner, respectively, on the (001) surface. The Fermi arcs connecting them traverse a distance of ~1 Å$^{-1}$, which is more than ten times larger than those in the Weyl semimetals TaAs (refs. 19-21) and MoTe$_2$ (refs. 22,23). Moreover, the Weyl points can shift with external conditions, such as the SOC strength, lattice constants, and atom positions etc, and even merge together, leading to annihilation of Weyl nodes. By contrast, the spin-1 and charge-2 points in CoSi are enforced at the high-symmetry momenta Γ and R by the little group symmetries with three- and four-dimensional irreducible representations, respectively[7,8].

Since the Co and Si atoms are strongly bonded by multiple covalent bonds in three dimensions, it is almost infeasible to obtain atomically flat surfaces by cleaving the single crystals. We thus tried to polish the surfaces of single crystals, and then repeatedly sputter and anneal the polished surfaces in vacuum. Eventually, we attained atomically flat (111) and (001) surfaces, as manifested by clear RHEED patterns in the insets of Fig. 1g,h.

We observed clear band dispersions in the ARPES measurements on these surfaces. The ARPES data collected on the (111) surface are summarized in Fig. 2. By varying the photon energy (*hv*) from 325 to 580 eV, we obtain the band dispersions along Γ-R, which is normal to the (111) surface, as shown in Fig. 2d. We observed two hole-like bands around Γ in the energy range within 0.5 eV below $E_F$. The two bands degenerate with an electron-like band at R. We further investigated the in-plane band structure on the (111) surface with *hv* = 435 eV, which measures the momentum plane containing the R and X points, as illustrated in Fig. 2a,b. We observed a circular FS patch at R in Fig. 2c, which arises from the electron-like band shown in Fig. 2e,f. The electron-like band degenerates with a relatively flat band at ~0.2 eV below $E_F$ at R in Fig. 2e,f. All the observations are consistent with the calculated bulk band structures, except that the splitting of the electron-like band along Γ-R is not resolved.

Next we focus on the APRES results taken on the (001) surface of CoSi. In the ARPES experiments on the (001) surface, we used soft X-rays and vacuum ultraviolet (VUV) lights, which can selectively probe the bulk and surface states. The ARPES data collected with soft X-rays are summarized in Fig. 3. We plot in Fig. 3d,e the FSs measured with $hv$ = 355 and 300 eV, respectively. The experimental FS at the BZ center in Fig. 3d is consistent with the calculated hole-like FS at Γ in Fig. 3b, while the one at the BZ corner in Fig. 3e is consistent with the calculated electron-like FS at R in Fig. 3c. This indicates that the measured momenta in Fig. 3d,e, correspond to the $k_z$ = 0 and π planes, respectively. Figure 3f,g shows the experimental band dispersions measured with $hv$ = 355 eV, which are well consistent with the calculated bands along Γ-M.

The excellent consistency in experiment and calculation provides solid evidence for the presence of spin-1 and charge-2 nodes in the bulk of CoSi. Since these degenerate points have nonzero Chern number, it is expected that surface Fermi arcs emanate from their projections on certain surfaces. On the (111) surface, the spin-1 point at Γ and charge-2 point at R are projected onto the same point at the surface BZ center, at which their Chern numbers cancel exactly. There are no topologically protected Fermi arcs on the (111) surface. By contrast, on the (001) surface, the Γ and R points are projected onto the surface BZ center $\bar{\Gamma}$ and corner $\bar{M}$, respectively. It is thus expected that two Fermi arcs connect the $\bar{\Gamma}$ and $\bar{M}$ points on the (001) surface.

In Fig. 3d,e, we already observed some extra features in the soft X-ray ARPES data, as compared with the calculated bulk FSs. These features may come from the (001) surface states. To confirm this, we have carried out APRES experiments on the (001) surface using VUV lights. The photoelectrons excited by VUV lights have a shorter escape depth as compared with those with soft X-ray. So the surface states could be detected more clearly in VUV ARPES experiments.

The ARPES data collected with VUV lights on the (001) surface are summarized in Fig. 4. The FSs measured with VUV lights in Fig. 4a,b are distinct from the calculated bulk FSs. The most remarkable feature in the experimental FSs is the Fermi arcs that connect the bulk FS pockets at $\bar{\Gamma}$ and $\bar{M}$. In Fig. 4g, we plot the Fermi arcs extracted from the ARPES intensity maps at three different constant energies in Fig. 4b-d. The two Fermi arcs are related by a π rotation about $\bar{\Gamma}$, which is constrained by time-reversal symmetry as the lattice symmetries in the bulk are broken on the (001) surface. The observed Fermi arcs are qualitatively consistent with the calculations[10].

To illuminate topological attributes of the Fermi arcs, we plot the band dispersions along three cuts #1, #2, and #3 in Fig. 4e, and compare them with the calculated bulk states projected onto the (001) surface in Fig. 4f. As the bulk bands #2 and #3 contact only at Γ and R, they are gapped in the $k_x$-$k_z$ planes with $k_y \neq 0$ and $\pi/a$, where the Chern number $C$ can be defined. We found that $C = +1$ (-1) for the $k_x$-$k_z$ planes with $0 < k_y < \pi/a$ ($0 > k_y > -\pi/a$). There should be one surface band that traverses the bulk band gap on the (001) projection line of each $k_x$-$k_z$ plane with $k_y \neq 0$ and $\pi/a$, as seen in Fig. 4e,f. The isoenergy contours of the surface bands are the Fermi arcs. Figure 4f shows that along the cuts #2 and #3, the surface bands bend back and then turn up on the way, crossing $E_F$ three times. When lowering the energy to -0.08 eV, the surface bands pass through the energy one time, resulting in the deformation of the Fermi arcs for different constant energies.

Although the Fermi arcs obviously deform with varying the constant energy, it does not destroy the connection between $\bar{\Gamma}$ and $\bar{M}$, confirming that the Fermi arcs in CoSi are noncontractible. This is essentially distinct from the Fermi arcs observed in Na$_3$Bi (ref. 28) and WC (ref. 27). While pairs of Fermi arcs connect the surface projections of Dirac points in Na$_3$Bi and triple points in WC at the energy of the bulk nodes, the Fermi arcs continuously deform into closed Fermi surfaces separated from the bulk states or even disappear with varying the constant energy[29,30]. The unprotected Fermi arcs indicate that the normal Dirac nodes and triply degenerate nodes have no chirality. By contrast, the noncontractible Fermi arcs observed in CoSi indicate that the three-component spin-1 node and charge-2 Dirac node are chiral with opposite Chern numbers $C = \pm 2$. Our ARPES experiments thus unambiguously prove the presence of chiral topological nodes beyond the Weyl nodes in condensed matter systems.

# References

[1] Wan, X., Turner, A. M., Vishwanath, A. & Savrasov, S. Y. Topological semimetal and Fermi-arc surface states in the electronic structure of pyrochlore iridates. *Phys. Rev. B* **83**, 205101 (2011).

[2] Weng, H., Fang, C., Fang, Z., Bernevig, B. A. & Dai, X. Weyl semimetal phase in noncentrosymmetric transition-metal monophosphides. *Phys. Rev. X* **5**, 011029 (2015).

[3] Huang, S. M. *et al*. A Weyl Fermion semimetal with surface Fermi arcs in the transition metal monopnictide TaAs class. *Nat. Commun.* **6**, 7373 (2015).

[4] Soluyanov, A. A. *et al*. Type-II Weyl semimetals. *Nature* **527**, 495–498 (2015).

[5] Xu, G., Weng, H., Wang, Z., Dai, X. & Fang, Z. Chern semimetal and the quantized anomalous Hall effect in $HgCr_2Se_4$. *Phys. Rev. Lett.* **107**, 186806 (2011).

[6] Bradlyn, B. *et al*. Beyond Dirac and Weyl fermions: unconventional quasiparticles in conventional crystals. *Science* **353**, aaf5037 (2016).

[7] Zhang, T. *et al*. Double-Weyl phonons in transition-metal monosilicides. *Phys. Rev. Lett.* **120**, 016401 (2018).

[8] Tang, P., Zhou, Q. & Zhang, S.-C. Multiple types of topological fermions in transition metal silicides. *Phys. Rev. Lett.* **119**, 206402 (2017).

[9] Chang, G. *et al*. Unconventional chiral fermions and large topological Fermi arcs in RhSi. *Phys. Rev. Lett.* **119**, 206401 (2017).

[10] Pshenay-Severin, D. A., Ivanov, Y. V., Burkov, A. A. & Burkov, A. T. Band structure and unconventional electronic topology of CoSi. Preprint at https://arXiv.org/abs/1710.07359 (2017).

[11] Wang, Z. *et al*. Dirac semimetal and topological phase transitions in $A_3Bi$ (A = Na, K, Rb). *Phys. Rev. B* **85**, 195320 (2012).

[12] Young, S. M. *et al*. Dirac semimetal in three dimensions. *Phys. Rev. Lett.* **108**, 140405 (2012).

[13] Wang, Z., Weng, H., Wu, Q., Dai, X. & Fang, Z. Three-dimensional Dirac semimetal and quantum transport in $Cd_3As_2$. *Phys. Rev. B* **88**, 125427 (2013).

[14] Heikkilä, T. T. & Volovik, G. E. Nexus and Dirac lines in topological materials. *New J. Phys.* **17**, 093019 (2015).

[15] Wieder, B. J., Kim, Y., Rappe, A. M. & Kane, C. L. Double Dirac semimetals in three dimensions. *Phys. Rev. Lett.* **116**, 186402 (2016).

[16] Weng, H., Fang, C., Fang, Z. & Dai, X. Topological semimetals with triply degenerate nodal points in θ-phase tantalum nitride. *Phys. Rev. B* **93**, 241202 (2016).

[17] Zhu, Z., Winkler, G. W., Wu, Q. S., Li, J. & Soluyanov, A. A. Triple point topological metals. *Phys. Rev. X* **6**, 031003 (2016).

[18] Weng, H., Fang, C., Fang, Z. & Dai, X. Co-existence of Weyl fermion and massless triply degenerate nodal points. *Phys. Rev. B* **94**, 165201 (2016).

[19] Lv, B. Q. *et al*. Experimental discovery of Weyl semimetal TaAs. *Phys. Rev. X* **5**, 031013 (2015).

[20] Xu, S.-Y. *et al*. Discovery of a Weyl fermion semimetal and topological Fermi arcs. *Science* **349**, 613–617 (2015).

[21] Lv, B. Q. *et al*. Observation of Weyl nodes in TaAs. *Nat. Phys.* **11**, 724–727 (2015).

[22] Deng, K. *et al*. Experimental observation of topological Fermi arcs in type-II Weyl semimetal $MoTe_2$. *Nat. Phys.* **12**, 1105 (2016).

[23] Huang, L. *et al*. Spectroscopic evidence for a type II Weyl semimetallic state in $MoTe_2$. *Nat. Mater.* **15**, 1155 (2016).

[24] Liu, Z. K. *et al*. Discovery of a three-dimensional topological Dirac semimetal, $Na_3Bi$. *Science* **343**, 864–867 (2014).

[25] Liu, Z. K. *et al*. A stable three-dimensional topological Dirac semimetal $Cd_3As_2$. *Nat. Mater.* **13**, 677–681 (2014).

[26] Lv, B. Q. *et al*. Observation of three-component fermions in the topological semimetal molybdenum phosphide. *Nature* **546**, 627–631 (2017).

[27] Ma, J.-Z. *et al*. Three-component fermions with surface Fermi arcs in tungsten carbide. *Nat. Phys.* **14**, 349-354 (2018).

[28] Xu, S.-Y. *et al*. Observation of Fermi arc surface states in a topological metal. *Science* **347**, 294–298 (2015).

[29] Kargarian, M., Randeria, M. & Lu, Y.-M. Are the surface Fermi arcs in Dirac semimetals topologically protected? *Proc. Natl Acad. Sci. USA* **113**, 8648–8652 (2016).

[30] Fang, C., Lu, L., Liu, J. & Fu, L. Topological semimetals with helicoid surface states. *Nat. Phys.* **12**, 936–941 (2016).

## METHODS

**Sample synthesis.** Single crystals of CoSi were grown by chemical vapor transport method. Co and Si powders in 1:1 molar ratio were put into a silica tube with the length of 200 mm and the inner diameter of 14 mm. Then, 200mg $I_2$ was added into tube as a transport reagent. The tube was evacuated down to $10^{-2}$ Pa and sealed under vacuum. The tubes were placed in two-zone horizontal tube furnace and the source and growth zones were raised to 950 K and 800 K in 2 days, and then held there for 7 days. The shiny crystals with lateral dimensions up to several millimeters can be obtained.

**Angle-resolved photoemission spectroscopy.** ARPES measurements were performed at the 'Dreamline' beamline of the Shanghai Synchrotron Radiation Facility (SSRF) with a Scienta Omicron DA30L analyzer. To obtain atomically flat surfaces for ARPES measurements, we polished the (111) and (001) surfaces of single crystals, and then repeatedly sputtered the surfaces and annealed the samples until clear RHEED patterns appeared.

**Band structure calculations**. First-principles calculations were performed by density functional theory (DFT) (ref. 31) within the Perdew-Burke-Ernzerhof (PBE) type exchange-correlation[32], and implemented in the Vienna *ab initio* simulation package (VASP) (ref. 33). A 20×20×20 k-mesh was used in the BZ for the self-consistent calculations, and all the calculations were in the absence of SOC. The three-dimensional FS calculations were employed by the tight-binding model of the bulk CoSi, which was obtained from maximally localized Wannier functions[34]. The experimental values of the atomic sites and lattice constant $a$ = 4.445 Å (ref. 35) were used in our calculations.

[31] Hohenberg, P. & Kohn, W. Inhomogeneous electron gas. Phys. Rev. 136, B864 (1964).

[32] Perdew, J. P., Burke, K. & Ernzerhof, M. Generalized gradient approximation made simple. *Phys. Rev. Lett*. **77**, 3865 (1996).

[33] Kresse, G. & Furthmüller, J. Efficient iterative schemes for *ab initio* total-energy calculations using a plane-wave basis set. *Phys. Rev. B* **54**, 11169–11186 (1996).

[34] Marzari, N. & Vanderbilt, D. Maximally localized generalized Wannier functions for composite energy bands. *Phys. Rev. B* **56**, 12847–12865 (1997).

[35] https://materials.springer.com/isp/crystallographic/docs/sd_0378292

**Acknowledgements**

We acknowledge Yigui Zhong and Jianyu Guan for assistance with RHEED measurements. This work was supported by the Ministry of Science and Technology of China (2016YFA0401000, 2016YFA0300600, 2016YFA0302400, 2016YFA0300504, 2015CB921300, and 2017YFA0302901), the National Natural Science Foundation of China (11622435, 11474340, 11422428, 11674369, 11474330, 11574394, 11774423, and 11774399), the Chinese Academy of Sciences (QYZDB-SSW-SLH043, XDB07000000, and XDPB08-1), and the Beijing Municipal Science and Technology Commission (No. Z171100002017018). Y.B.H. acknowledges support by the CAS Pioneer "Hundred Talents Program" (type C).

**Author contributions**

T.Q. and Y.-J.S. supervised the project. Z.-C.R., H.L. and T.Q. performed ARPES measurements with the assistance of B.-B.F., W.-H.F., J.-J.L. and Y.-B.H.; Z.-C.R., C.-Y.T. and Y.-J.S. processed the sample surfaces with the assistance of Z.-H.L. and Y.-W.L.; T.-T.Z. and H.-M.W. performed *ab initio* calculations; S.-J.T., C.-H.L., H.-C.L., L.W., Y.-G.S. and Z.-L.L. synthesized the single crystals; Z.-C.R, H.L., T.Q. and Y.-J.S. analysed the experimental data; Z.-C.R., T.-T.Z., T.Q. and Y.-J.S. plotted the figures; T.Q., C.F. and H.D. wrote the manuscript.

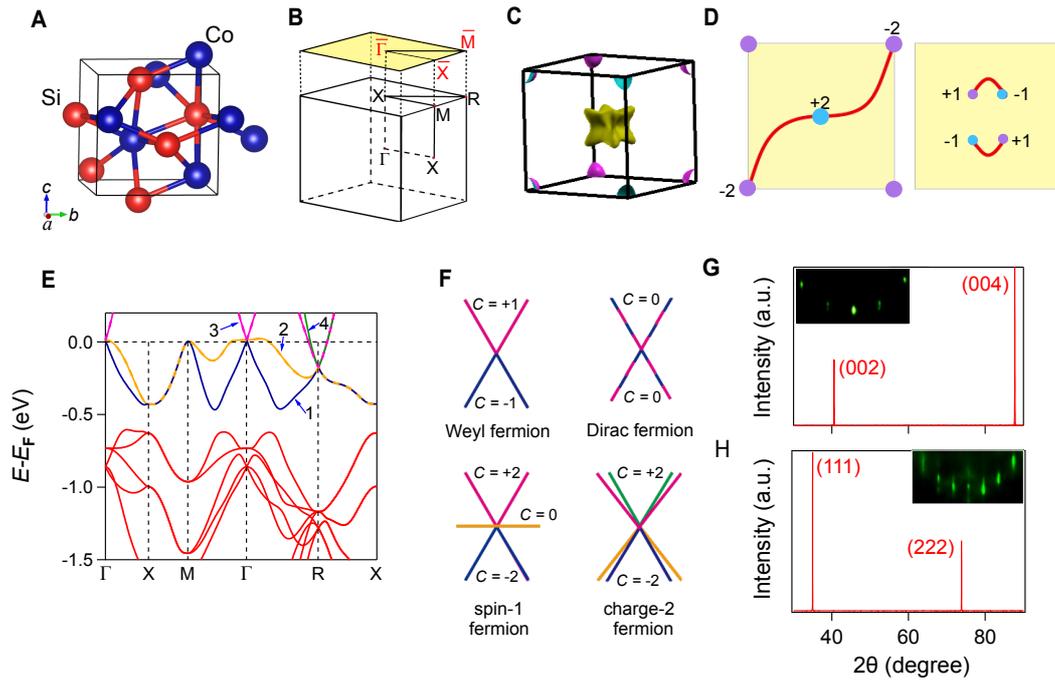

**Figure 1 | Crystal stricture and calculated electronic structure of CoSi. a**, Crystal structure of CoSi. **b**, Bulk BZ and (001) surface BZ of CoSi. **c**, Calculated FSs in the bulk BZ. **d**, Schematics of the Fermi arcs connecting the projections of two nodes with opposite chiral charges for CoSi (left) and Weyl semimetals (right). **e**, Calculated bulk band structure along high-symmetry lines without SOC. **f**, Schematics of the band structures of Weyl node, Dirac node, chiral spin-1 node, and charge-2 Dirac node. **g**, XRD patterns measured on the (001) plane of CoSi. The inset shows the RHEED pattern of CoSi after repeatedly sputtered and annealed. **h**, Same as **g** but measured on the (111) surface.

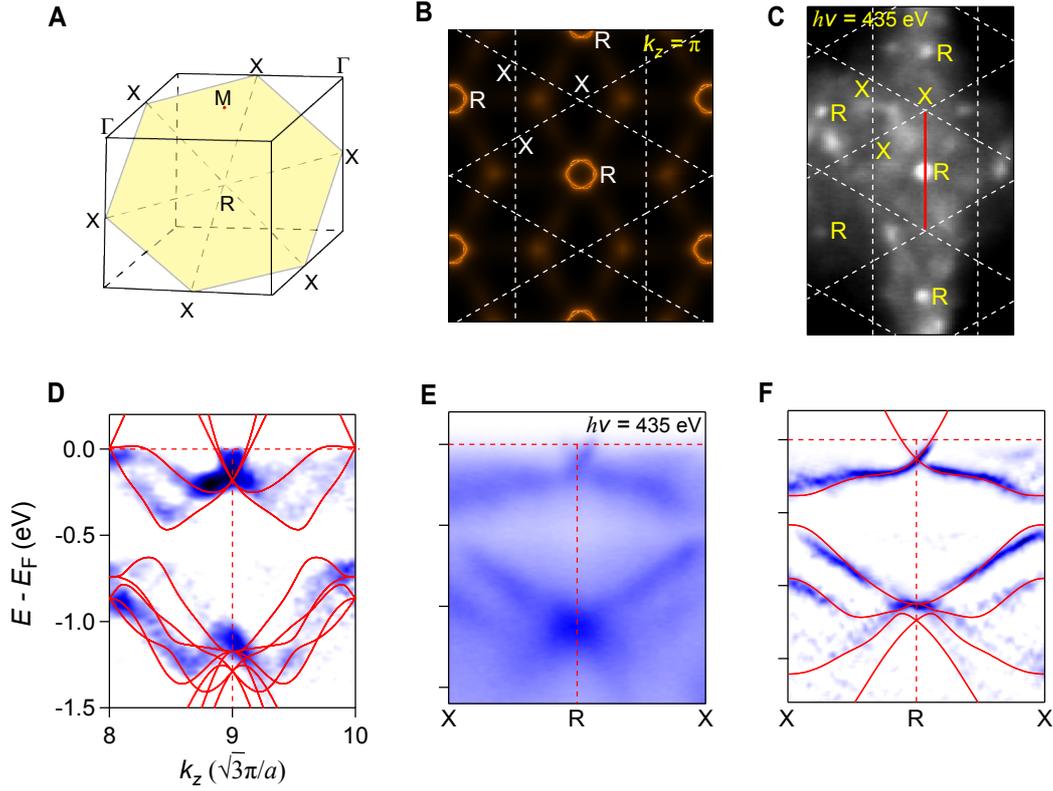

**Figure 2 | FSs and band dispersions measured on the (111) surface of CoSi. a**, Hexagon in the 3D BZ indicates the (111) plane that contains the R and X points. **b,c**, Calculated (**b**) and experimental (**c**) intensity plots at $E_F$, showing the FSs in the (111) plane indicated in **a**. Solid lines represent the intersecting lines of the (111) plane and bulk BZ boundary. Dashed lines represent the (111) surface BZ. **d**, Curvature intensity plot of the ARPES data measured with varying $h\nu$ from 325 to 580 eV, showing the band dispersions along R-Γ-R. The inner potential was set to be 27 eV to fit to the periodicity. **e,f**, Intensity and curvature intensity plots of the ARPES data along X-R-X, respectively. For comparison, we plot the calculated bands along R-Γ-R and X-R-X as red lines on top of the experimental data in **d** and **f**, respectively. The ARPES data in **c**, **e**, and **f** were collected with $h\nu = 435$ eV.

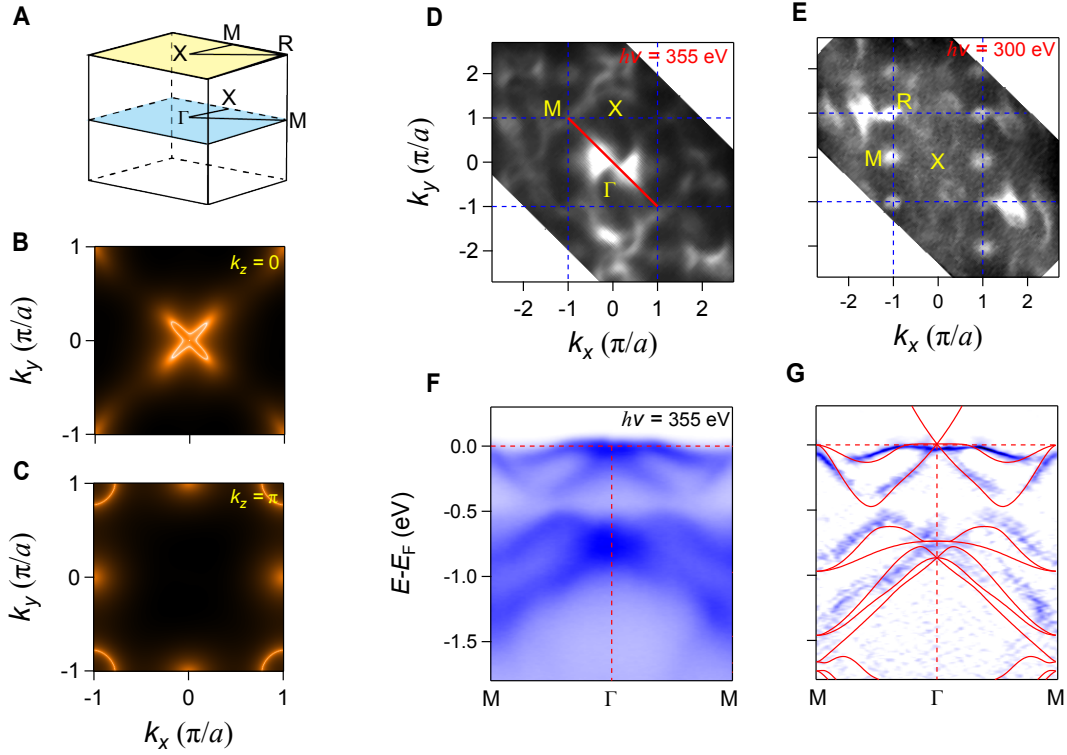

**Figure 3 | FSs and band dispersions measured with soft X-rays on the (001) surface of CoSi. a**, 3D bulk BZ with the $k_z = 0$ and $\pi/a$ planes indicated by colors. **b,c**, Calculated bulk FSs in the $k_z = 0$ and $\pi/a$ planes, respectively. **d,e**, ARPES intensity maps at $E_F$ measured with $h\nu = 355$ and 300 eV, respectively. **f,g**, Intensity and curvature intensity plots of the ARPES data along M-Γ-M collected with $h\nu = 355$ eV, respectively. For comparison, we plot the calculated bands along M-Γ-M as red lines on top of the experimental data in **g**.

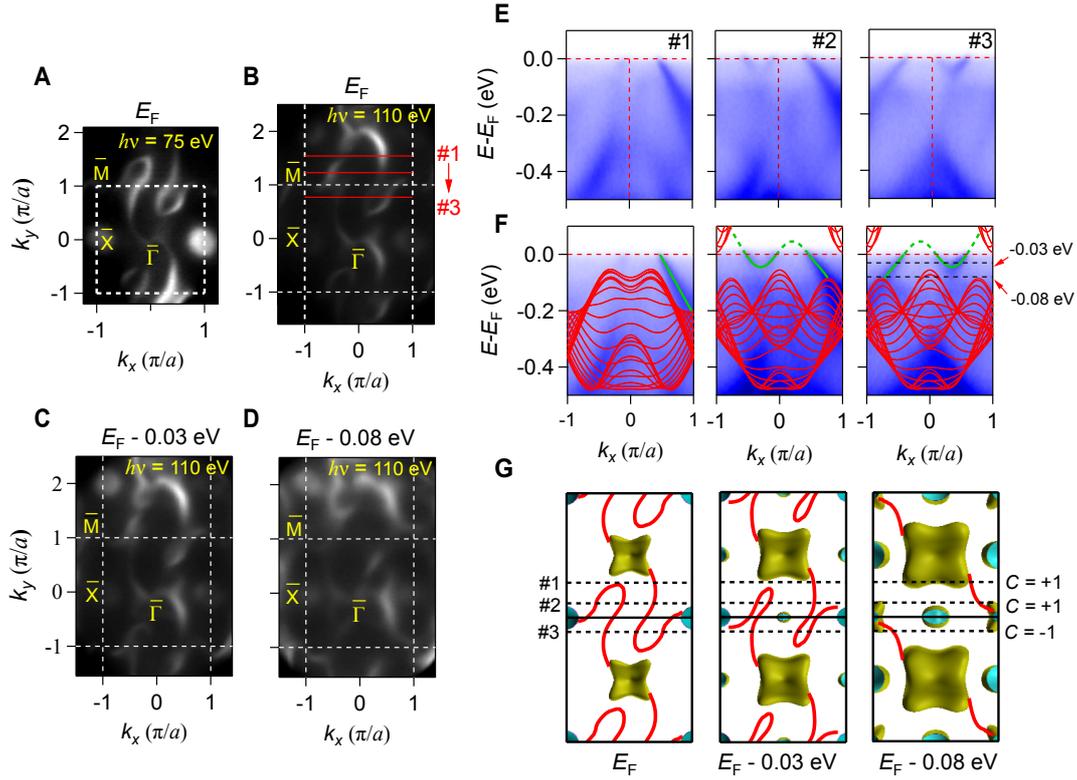

**Figure 4 | Surface Fermi arcs measured with VUV lights on the (001) surface of CoSi. a**, ARPES intensity map at $E_F$ measured with $h\nu$ = 75 eV. **b-d**, ARPES intensity maps at $E_F$, $E_F$-0.03 eV, $E_F$-0.08 eV measured with $h\nu$ = 110 eV, respectively. **e**, ARPES intensity plots showing the band dispersions along the cuts #1, #2, and #3, whose momentum locations are indicated in **b**. For comparison, we plot the projections of the calculated bulk bands as red lines on top of the experimental data in **f**. Solid and dashed green lines in **f** are guides to eyes for the surface state bands. **g**, Surface Fermi arcs extracted from the ARPES intensity maps at $E_F$ (left), $E_F$-0.03 eV (middle), and $E_F$-0.08 eV (right), showing the connection between the (001) surface projections of bulk FSs at $\bar{\Gamma}$ and $\bar{M}$. The Chern number $C$ = +1 for the cuts #1 and #2 and $C$ = -1 for the cut #3.